\documentclass[aps,prl,twocolumn,groupedaddress,showpacs]{revtex4}
\usepackage{graphicx,epsfig}
\usepackage[usenames]{color}
\usepackage{amsmath}
\date{\today}

\newcommand{\bi}{{\bf i}}
\newcommand{\bj}{{\bf j}}
\newcommand{\bk}{{\bf k}}

\newcommand{\cZ}{{\mathcal Z}}

\begin{document}
\title{Fractional exclusion statistics -- the method to describe interacting particle systems as ideal gases}
\author{Drago\c s-Victor Anghel}
\affiliation{Department of Theoretical Physics, Horia Hulubei National Institute for Physics and Nuclear Engineering, 30 Reactorului street, P.O.BOX MG-6, M\u agurele, Jud. Ilfov, Romania}
%
%\pacs{05.30.-d}{Quantum statistical mechanics}
%\pacs{05.30.Ch}{Quantum ensemble theory}
%\pacs{05.30.Pr}{Fractional statistics systems (anyons, etc.)}
\begin{abstract}

I show that if the total energy of a system of interacting particles may be written as a sum of quasiparticle energies, then the system of quasiparticles can be viewed in general as an ideal gas with fractional exclusion statistics (FES). The general method for calculating the FES parameters is also provided. 

The interacting particle system cannot be described as an ideal gas of Bose and Fermi quasiparticles except in trivial situations. 

\end{abstract}

\maketitle

\section{Introduction}\label{intro}

A fractional exclusion statistics (FES) system \cite{PhysRevLett.67.937.1991.Haldane,PhysRevLett.73.922.1994.Wu,PhysRevLett.73.2150.1994.Isakov} consists of a countable number of species. I will count the species using the indexes $i$ and $j$. Each species contains $G_i$ single-particle states and $N_i$ particles. The FES character of the systems consists in the dependence of the $G_i$'s on the $N_i$'s. For example by adding $\delta N_i$ particles in the species $i$ the dimensions of all the other species change according to $\delta G_j=-\alpha_{ji}\delta N_i$, for any $i$ and $j$. The parameters $\alpha_{ij}$ are called the \textit{FES parameters}. 

In practice, the species may be different types of particles that coexist in the same system, quasiparticle excitations in the lowest Landau level in the fractional quantum Hall effect \cite{PhysRevLett.67.937.1991.Haldane,PhysRevLett.73.922.1994.Wu}, excitations \cite{PhysRevLett.67.937.1991.Haldane} or motifs of spins in spin chains \cite{PhysRevE.84.021136.2011.Liu,PhysRevE.85.011144.2012.Liu}, elementary volumes obtained by coarse-graining in the phase-space of a system \cite{PhysRevLett.73.2150.1994.Isakov,JPhysA.40.F1013.2007.Anghel,PhysLettA.372.5745.2008.Anghel,EPL.90.10006.2010.Anghel}, etc. 
%
%The thermodynamics and statistical mechanics of systems was calculated mainly by Wu \cite{PhysRevLett.73.922.1994.Wu} and Isakov \cite{PhysRevLett.73.2150.1994.Isakov}, whereas 
Some amendments to the FES formalism have been introduced more recently, in Refs. \cite{JPhysA.40.F1013.2007.Anghel,EPL.87.60009.2009.Anghel,PhysRevLett.104.198901.2010.Anghel,PhysRevLett.104.198902.2010.Wu}. 

In this paper I shall use the ansatz 
\begin{equation}
\alpha_{ij}=G_i a_{ij} +a_{i}\delta_{ij}. \label{alpha_nd}
\end{equation}
which applies quite generally in quasi-continuous systems \cite{EPL.90.10006.2010.Anghel}.

The equilibrium distribution of particles in the species may be calculated in two equivalent formulations: I call them the Bose and Fermi perspectives \cite{EPL.90.10006.2010.Anghel}. 

In the Bose perspective I use $G_i$ to denote the number of available single-particle states of species $i$. In such a case, the number of microconfigurations in which we can find the system is $W=\prod_i\{(G_i+N_i-1)!/[N_i!(G_i-1)!]\}$. If we add a small perturbation, $\{\delta N_i\}$, to the particle distribution, $W$ becomes
\begin{equation}
W_{B} = \prod_{i} \frac{[G_{i}+N_{i}-1+(1-a_i)\delta N_i -\sum_{j}\alpha_{ij} \delta N_{j}]!} {(N_{i}+\delta N_{i})!(G_{i}-1-a_i\delta N_i-\sum_{j} \alpha_{ij}\delta N_{j})!} . \label{W_ext_pert}
\end{equation}
Associating the energies $\{\epsilon_i\}$ to the particles in each of the species, we write the partition function, $\cZ$ \cite{EPL.90.10006.2010.Anghel}, which we then maximize to obtain a set of equations for the equilibrium particle distribution, 
\begin{equation}
\beta(\mu-\epsilon_i)+\ln\frac{[1+b_i]^{1-a_{i}}}{b_i} = \sum_{j} G_{j} a_{ji} \ln[1+b_{j}] .  \label{inteq_for_n1}
\end{equation}
where $\beta=1/(k_BT)$, $\mu$ is the chemical potential of the system, and $b_i=N_i/G_i$ is the equilibrium particle population. 

In the Fermi perspective, $T_i\equiv G_i+N_i-1$, and $W_F=\prod_i\{T_i!/[N_i!(T_i-N_i)!]\}$. At the variations $\{\delta N_i\}$, we obtain
\begin{eqnarray}
  W_{F} &=& \prod_{i}\left\{\frac{(T_{i}-a_i\delta N_{i} -G_{i}\sum_{j}a_{ij}\delta N_{j})!}{[T_{i}-N_{i}-(1+a_i) \delta N_{i}-T_{i}\sum_{j}a_{ij}\delta N_{j}]!} \right. \nonumber \\
  && \left.\times\frac{1}{(N_{i}+\delta N_{i})!}\right\} , \label{W_ext_pert_f}
\end{eqnarray}
and the equations for the equilibrium particle populations, $f_i=N_i/T_i$, are \cite{EPL.90.10006.2010.Anghel}
\begin{equation}
  \beta(\mu_i-\epsilon_i)+\ln\frac{[1-f_i]^{1+a_{i}}}{f_i} = -\sum_{j} T_{j}a_{ji}\ln[1-f_{j}].  \label{inteq_for_n1f}
\end{equation}

I will show below that systems of interacting particles may be described as ideal FES systems. For interacting bosons, more natural is the Bose perspective, whereas for interacting fermions it is more convenient to employ the Fermi perspective. 

Equations (\ref{inteq_for_n1}) and (\ref{inteq_for_n1f}) can be readily transformed into integral equations in the quasicontinuous case. If instead of the index $i$ (or $j$) we introduce the quasicontinuous variable $\bi$ (or $\bj$) -- which may be a multidimensional variable, like the quasi-momentum, or a one-dimensional variable, like the quasi-energy -- of density of states (DOS) $\sigma(\bi)$, then Eqs. (\ref{inteq_for_n1}) and (\ref{inteq_for_n1f}) become \cite{EPL.90.10006.2010.Anghel}
\begin{equation}
\beta(\mu_\bi-\epsilon_\bi)+\ln\frac{[1+b_\bi]^{1-a_{\bi}}}{b_\bi}=\int \sigma(\bj)\ln[1+b_{\bj}]a_{\bj\bi}\,d\bj .
\label{inteq_for_n2c}
\end{equation}
and
\begin{equation}
\beta(\mu_\bi-\epsilon_\bi)+\ln\frac{[1-f_\bi]^{1-a_{\bi}}}{f_\bi} 
=-\int \sigma(\bj)\ln[1-f_{\bj}]a_{\bj\bi}\,d\bj, \label{inteq_for_n2f}
\end{equation}
respectively. 

% The thermodynamics of different FES systems have been calculated by several authors (see  for example Refs. \cite{PhysRevLett.73.922.1994.Wu,PhysRevLett.73.2150.1994.Isakov,PhysRevLett.72.600.1994.Veigy,PhysLettA212.299.1996.Isakov,NewDevIntSys.1995.Bernard,PhysRevLett.73.3331.1994.Murthy,PhysRevB.60.6517.1999.Murthy,PhysRevE.75.61120.2007.Potter,PhysRevE.76.61112.2007.Potter,EPL.90.10006.2010.Anghel,EPL.94.60004.2011.Anghel,JPA35.7255.2002.Anghel,PhysRevB.56.4422.1997.Sutherland,PhysRevLett.85.2781.2000.Iguchi}).

\section{The quasiparticles}

\subsection{The ideal gas description}

Suppose we have a system of interacting particles which we want to describe as an ideal gas of quasiparticles. For this we introduce the quasiparticle energies, $\{\tilde\epsilon_i\}$, which we want to satisfy three conditions, specific to ideal gases:
\begin{subequations} \label{conditions}
\begin{eqnarray}
  &\bullet& E(\{n_i\}) = \sum_i n_i\tilde\epsilon_i, \label{cond_entot} \\
  &\bullet& \text{the energies }\{\tilde\epsilon_i\} \text{ are well defined and therefore} \nonumber \\
  && \text{independent of the set of occupation numbers,} \nonumber \\ 
  && \{n_i\}; \label{cond_tepsi} \\
  &\bullet& \text{the equilibrium populations, }\langle n_i\rangle(T,\mu,\tilde\epsilon_i), \text{ are} \nonumber \\ 
  && \text{functions of only } T, \mu \text{ and }\tilde\epsilon_\bi \text{ and are independent}\nonumber \\
  && \text{of the populations.}\label{cond_n}
\end{eqnarray}
\end{subequations}

The heat capacity of any system (in units of $k_B$) is
\begin{equation}
  \frac{C_V}{k_B} \equiv \left(\frac{\partial U}{\partial T}\right)_N = \left(\frac{\partial U}{\partial T}\right)_\mu - \left(\frac{\partial U}{\partial\mu}\right)_T\left(\frac{\partial N}{\partial T}\right)_\mu\left(\frac{\partial N}{\partial\mu}\right)_T^{-1}. \label{CV_ideal}
\end{equation}
If the conditions (\ref{conditions}) are satisfied, then
\begin{subequations}\label{der_part_CV}
\begin{eqnarray}
  \left(\frac{\partial U}{\partial(k_B T)}\right)_\mu &=& \sum_i \tilde\epsilon_i\left(\frac{\partial \langle n_i\rangle}{\partial(k_B T)}\right)_\mu , \label{partE_partT} \\
  \left(\frac{\partial U}{\partial\mu}\right)_T &=& \sum_i \tilde\epsilon_i\left(\frac{\partial \langle n_i\rangle}{\partial\mu}\right)_T , \label{partE_partmu} \\
  \left(\frac{\partial N}{\partial(k_B T)}\right)_\mu &=& \sum_i \left(\frac{\partial \langle n_i\rangle}{\partial(k_B T)}\right)_\mu , \label{partN_partT} \\
  \left(\frac{\partial N}{\partial\mu}\right)_T &=& \sum_i \left(\frac{\partial\langle n_i\rangle}{\partial\mu}\right)_T , \label{partN_partmu}
\end{eqnarray}
\end{subequations}
where we used the notation $U(T,\mu)\equiv \langle E\rangle_{T,\mu}$ for the internal energy of the system, at temperature $T$ and chemical potential, $\mu$. 

%\subsection{The thermodynamics of the ideal Fermi gas}

For an ideal gas of fermions in the quasicontinuous limit and with a density of states (DOS) of the form $\tilde\sigma(\tilde\epsilon)\equiv C\tilde\epsilon^s$, with $C$ and $s$ being two constants \cite{PhysRevA.35.4354.1987.Bagnato,PhysRevA.44.7439.1991.Bagnato}, the total particle number and the internal energy are 
\begin{subequations}\label{NUC_id_cont}
\begin{eqnarray}
  N &=& C (k_BT)^{s+1}\Gamma(s+1)Li_{s+1}\left(-e^{\beta\mu}\right)\ \text{and} \label{N_id_cont} \\
  U &=& C (k_BT)^{s+2}\Gamma(s+2)Li_{s+2}\left(-e^{\beta\mu}\right), \label{U_id_cont} 
\end{eqnarray}
respectively, where the function $Li_n$ is the polylogarithmic function of order $n$ \cite{Lewin:book,ActaPhysicaPolonicaB.40.1279.2009.Lee}. 

Using Eqs. (\ref{der_part_CV}) we calculate the specific heat (also in units of $k_B$):
\begin{eqnarray}
  \frac{c_V}{k_B}\equiv\frac{C_V}{k_BN} &=& \left[(s+1)(s+2) \frac{Li_{s+2}\left(-e^{\beta\mu}\right)}{Li_{s+1}\left(-e^{\beta\mu}\right)}\right. \nonumber \\
  && \left.-(s+1)^2 \frac{Li_{s+1}\left(-e^{\beta\mu}\right)}{Li_{s}\left(-e^{\beta\mu}\right)}\right]. \label{c_V_id_cont}
\end{eqnarray}
\end{subequations}
In the low temperature limit $\beta\mu\gg1$ and in the lowest orders of approximation, Eqs. (\ref{NUC_id_cont}) become 
\begin{subequations}\label{NUC_id_lowT}
\begin{eqnarray}
  N &=& C\mu^{s+1}/(s+1)\equiv C\tilde\epsilon_F^{s+1}/(s+1), \label{N_id_lowT} \\
  U &=& C\mu^{s+2}/(s+2), \label{U_id_lowT} \\
  \frac{c_V}{k_B} &\approx& (s+1)\frac{\pi^2}{3}\frac{k_BT}{\mu}\approx (s+1)\frac{\pi^2}{3}\frac{k_BT}{\tilde\epsilon_F}, \label{c_V_id_lowT}
\end{eqnarray}
\end{subequations}
where Eq. (\ref{N_id_lowT}) defines the Fermi energy, $\tilde\epsilon_F=[(s+1)N/C]^{1/(s+1)}$, and Eq. (\ref{c_V_id_lowT}) may be put into the low temperature universal form \cite{Stone:book,EPL.94.60004.2011.Anghel},
\begin{equation}
  \frac{C_V}{k_B} \equiv \frac{Nc_V}{k_B} = \frac{\pi^2}{3}k_BT\tilde\sigma(\tilde\epsilon_F). \label{CV_univ}
\end{equation}

\subsection{The quasiparticle gas}

The gases of quasiparticles used to describe systems of interacting particles do not satisfy in general the conditions (\ref{conditions}). An example shown in Ref. \cite{arXiv:1204.4064.Anghel} is the Landau's Fermi liquid theory (FLT). The quasiparticle energy, $\tilde\epsilon_i$, depends on the occupation of the other quasiparticle states, i.e. $\tilde\epsilon_i\equiv\tilde\epsilon_i(\{n_j\})$, where $\{n_j\}$ denotes the set of all occupation numbers. In such a case the condition (\ref{cond_tepsi}) is not satisfied and through $\tilde\epsilon_i$, the population $\langle n_i\rangle(T,\mu,\tilde\epsilon_i)$ depends on the populations of the other quasiparticle levels, violating the condition (\ref{cond_n}) as well. Moreover, in the FLT the sum of the energies of the quasiparticles is not equal to the total energy of the system, violating also condition (\ref{cond_entot}) (see \cite{arXiv:1204.4064.Anghel}).

Let us now see how the heat capacity of the system can be calculated. For this I will suppose that condition (\ref{cond_entot}) is true, otherwise the gas of quasiparticles may not be used for this purpose \cite{arXiv:1204.4064.Anghel}. 

If $\tilde\epsilon_i$ is a function of $\{n_j\}$, then Eqs. (\ref{partE_partT}) and (\ref{partE_partmu}) are not valid because $\tilde\epsilon_i$ varies with $T$ and $\mu$: 
\begin{subequations}\label{dteps_dTdmu}
\begin{eqnarray}
  \frac{\partial\tilde\epsilon_i(T,\mu)}{\partial(k_BT)} &=& \sum_j\frac{\partial\tilde\epsilon_i(T,\mu)}{\partial\langle n_j\rangle} \frac{\partial\langle n_j\rangle(T,\mu)}{\partial(k_BT)}, \label{dteps_dT} \\ 
  \frac{\partial\tilde\epsilon_i(T,\mu)}{\partial\mu} &=& \sum_j\frac{\partial\tilde\epsilon_i(T,\mu)}{\partial\langle n_j\rangle} \frac{\partial\langle n_j\rangle(T,\mu)}{\partial\mu}. \label{dteps_dmu}
\end{eqnarray}
\end{subequations}

On the other hand, the derivatives of the populations are given by the equations 
\begin{subequations} \label{dni_dTdmu}
\begin{equation}
  \frac{\partial\langle n_i\rangle(T,\mu)}{\partial(k_BT)} = \frac{\partial\langle n_i\rangle(T,\mu,\tilde\epsilon_i)}{\partial(k_BT)} + \frac{\partial\langle n_i\rangle(T,\mu,\tilde\epsilon_i)}{\partial\tilde\epsilon_i} \frac{\partial\tilde\epsilon_i(T,\mu)}{\partial(k_BT)} , \label{dni_dT}
\end{equation}
\begin{equation}
  \frac{\partial\langle n_i\rangle(T,\mu)}{\partial\mu} = \frac{\partial\langle n_i\rangle(T,\mu,\tilde\epsilon_i)}{\partial\mu} + \frac{\partial\langle n_i\rangle(T,\mu,\tilde\epsilon_i)}{\partial\tilde\epsilon_i} \frac{\partial\tilde\epsilon_i(T,\mu)}{\partial\mu} . \label{dni_dmu}
\end{equation}
\end{subequations}

Plugging Eqs. (\ref{dteps_dTdmu}) into (\ref{dni_dTdmu}) we obtain a self-consistent system of equations for the variation of the occupation numbers with $T$ and $\mu$. 

Using the solutions of Eqs. (\ref{dni_dTdmu}) together with Eqs. (\ref{dteps_dTdmu}), we can calculate $\partial U/\partial(k_BT)$, $\partial U/\partial\mu$, $\partial N/\partial(k_BT)$, and $\partial N/\partial\mu$, to finally obtain the heat capacity (\ref{CV_ideal}), where Eqs. (\ref{partE_partT}) and (\ref{partE_partmu}) become
\begin{subequations}\label{der_part_CV_qp}
\begin{eqnarray}
  \left(\frac{\partial U}{\partial(k_B T)}\right)_\mu &=& \sum_i \left[ \tilde\epsilon_i\left(\frac{\partial \langle n_i\rangle}{\partial(k_B T)}\right)_\mu \right. \nonumber \\
  && \left. + \langle n_i\rangle \left(\frac{\partial\tilde\epsilon_i}{\partial(k_B T)}\right)_\mu\right], \label{partE_partT_qp}
\end{eqnarray}
\begin{equation}
  \left(\frac{\partial U}{\partial\mu}\right)_T = \sum_i \left[ \tilde\epsilon_i\left(\frac{\partial \langle n_i\rangle}{\partial\mu}\right)_T + \langle n_i\rangle \left(\frac{\partial\tilde\epsilon_i}{\partial\mu}\right)_T \right], \label{partE_partmu_qp} 
\end{equation}
\end{subequations}

\section{The ideal FES gas}

Assuming that we choose the quasiparticle energies, $\tilde\epsilon_i$, in such a way that condition (\ref{cond_entot}) is satisfied, let us see how we can satisfy also the conditions (\ref{cond_tepsi}) and (\ref{cond_n}).

The principle of the method is given in Refs. \cite{PhysLettA.372.5745.2008.Anghel,PhysLettA.376.892.2012.Anghel,RJP.54.281.2009.Anghel}. 

%For concreteness, let's assume from now that the particles in the system are \textit{bosons}. It is easy to repeat the procedure for fermions. 

I change to the quasicontinuous description, $i\to\bi$, and I assume for simplicity that $\bi$ is a 1D quantity (e.g. the energy of the free particles, $\epsilon$ \cite{PhysLettA.372.5745.2008.Anghel,PhysLettA.376.892.2012.Anghel,RJP.54.281.2009.Anghel}). %In this way I establish a functional of the occupation numbers, 
The quasiparticle energy, $\tilde\epsilon_\bi(\{n_\bj\})$, is then a functional of the occupation numbers, $\{n_\bj\}$, or, for the equilibrium distribution, it is a functional of the populations -- $\tilde\epsilon_\bi(\{\langle n_\bj\rangle\})$. 

I choose $\bi$ or $\tilde\epsilon_\bi$ in such a way that if $\bi\le\bj$, then $\tilde\epsilon_\bi(\{\langle n_\bk\rangle\})\le\tilde\epsilon_\bj(\{\langle n_\bk\rangle\})$. In this way I establish a bijective correspondence between $\bi$ and $\tilde\epsilon$, which one may invert and write $\bi_{\tilde\epsilon}(\{\langle n_\bj\rangle\})$ or $\bi_{\tilde\epsilon}(\{\langle n_{\tilde\epsilon'}\rangle\})$. 

If $\sigma(\bi)$ is the DOS in the variable $i$, then 
\begin{equation}
  \tilde\sigma(\tilde\epsilon) = \sigma(\bi)\left|\frac{d\bi}{d\tilde\epsilon}\right| \equiv \sigma(\bi)\left|\frac{d\tilde\epsilon}{d\bi}\right|^{-1}; \label{def_til_sig}
\end{equation}
$\tilde\sigma(\tilde\epsilon)$ is a functional of $\{\langle n_\bj\rangle\}$. With the aid of $\sigma(\bi)$ and $\tilde\sigma(\tilde\epsilon)$ we define the particle densities, $\rho(\bi)=\sigma(\bi)\langle n_\bj\rangle$ and $\tilde\rho(\tilde\epsilon) =\tilde\sigma(\tilde\epsilon)\langle n_{\tilde\epsilon}\rangle$. 

Now I transform the quasiparticle gas into an ideal gas by simply changing the perspective: the usual perspective is to see the quantum numbers of the ideal gas, $\bi$, as fixed and the quasiparticle energies, $\tilde\epsilon_\bi(\{\langle n_\bj\rangle\})$, as functionals of the populations (Fig. \ref{FES_principle_qp} a). When we invert the perspective we see the quasiparticle energies, $\tilde\epsilon$, fixed and the free particle quantum numbers as functionals of the populations, $\bi_{\tilde\epsilon}(\{\langle n_{\tilde\epsilon}\rangle\})\equiv\bi_{\tilde\epsilon}[\rho(\tilde\epsilon')]$ (Fig. \ref{FES_principle_qp} b).

\begin{widetext}
\begin{center}
\begin{figure}[h]
  \includegraphics[width=15cm]{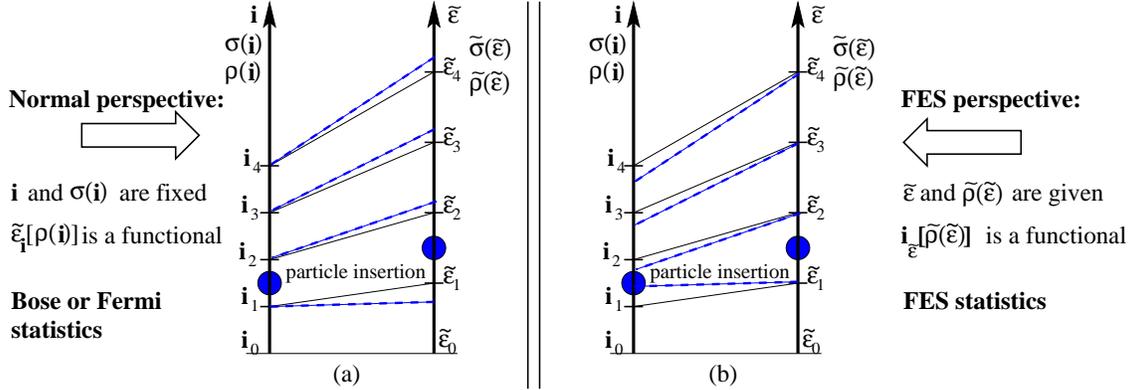}
  % FES_principle_qp.eps: 0x0 pixel, 300dpi, 0.00x0.00 cm, bb=0 0 1362 432
  \caption{(Color online) The two perspectives: (a) the the free-particle quantum numbers, $\bi$, are fixed and the quasiparticle energies $\tilde\epsilon_\bi[\rho(\bj)]$ are functionals of the particle density, $\rho(\bj)$; (b) the quasiparticle energies, $\tilde\epsilon$, are fixes and the free-particle quantum numbers, $\bi_{\tilde\epsilon}[\tilde\rho(\tilde\epsilon')]$, are functionals of the particle density, $\tilde\rho(\tilde\epsilon)$.}
  \label{FES_principle_qp}
\end{figure}
\end{center}
\end{widetext}

By the relation (\ref{def_til_sig}) and since $\sigma(\bi)$ is a quantity that is fixed by the properties of the single-particle states, $\tilde\sigma(\tilde\epsilon)$ becomes a functional of the particle density, $\tilde\rho(\tilde\epsilon)$. This property ensues FES. 

To show this I coarse-grain both axes, $\bi$ and $\tilde\epsilon$. Each interval, $[\bi_m,\bi_{m+1}]$ on the $\bi$ axis or $[\tilde\epsilon_m,\tilde\epsilon_{m+1}]$ on the $\tilde\epsilon$ axis, represents a species, with $G_m=\sigma(\bi_m)(\bi_{m+1}-\bi_m) = \tilde\sigma(\tilde\epsilon_m)(\tilde\epsilon_{m+1}-\tilde\epsilon_m)$ and $N_m=\rho(\bi_m)(\bi_{m+1}-\bi_m) = \tilde\rho(\tilde\epsilon_m)(\tilde\epsilon_{m+1}-\tilde\epsilon_m)$ (assuming that the intervals are small enough to use the linear approximation); I use the letters $m$ and $n$ to designate the species. 

In the FES perspective, the insertion of $\delta N_m$ particles in the species $m$ causes a change of the interval $[\bi_n,\bi_{n+1}]$ on the $\bi$ axis by
\begin{equation}
  \delta N_m\left[\frac{\delta\bi_{n+1}}{\delta\rho(\tilde\epsilon_m)} -\frac{\delta\bi_{n}}{\delta\rho(\tilde\epsilon_m)}\right] = \delta N_m(\bi_{m+1}-\bi_m) \frac{d}{d\bi}\left[\frac{\delta\bi_{n}}{\delta\rho(\tilde\epsilon_m)}\right], \label{der_fnc}
\end{equation}
where by $\delta\bi_{n}/\delta\rho(\tilde\epsilon_m)$ I denote the functional derivative, which I assume to be analytic. 

The change of the interval $[\bi_n,\bi_{n+1}]$ leads to the change of the number of states in the species $n$ by 
\begin{eqnarray}
  \delta G_n &=& \delta N_m\left[\frac{\delta\bi_{n+1}}{\delta\rho(\tilde\epsilon_m)}\sigma(\bi_{n+1})- \frac{\delta\bi_{n}}{\delta\rho(\tilde\epsilon_m)}\sigma(\bi_{n})\right] \nonumber \\
  &=& \delta N_m(\bi_{m+1}-\bi_m) \frac{d}{d\bi}\left[\frac{\delta\bi_{n}}{\delta\rho(\tilde\epsilon_m)}\sigma(\bi_{n})\right] \label{der_fnc_G_i}
\end{eqnarray}

Mapping back, $\bi$ onto $\tilde\epsilon$, I can express the change of the number of states in the species $n$ as
\begin{equation}
  \delta G_n = \delta N_m(\tilde\epsilon_{n+1}-\tilde\epsilon_n) \frac{d}{d\tilde\epsilon} \left[\frac{\delta\tilde\epsilon_{n}}{\delta\tilde\rho(\tilde\epsilon_m)} \tilde\sigma(\tilde\epsilon_{n})\right] \equiv -\alpha_{\tilde\epsilon_n\tilde\epsilon_m}\delta N_m, \label{der_fnc_G_e}
\end{equation}
where by $\delta\tilde\epsilon_{n}/\delta\rho(\tilde\epsilon_m)$ I denote the functional derivative of $\tilde\epsilon_{n}$ with respect to the variation of $\delta\rho(\tilde\epsilon_m)$, where $\bi$ is fixed -- this calculation trick does not change the noninteracting character of the quasiparticles.

The last part of Eq. (\ref{der_fnc_G_e}) gives us the FES parameter, $\alpha_{\tilde\epsilon_n\tilde\epsilon_m}$. We notice that $\alpha_{\tilde\epsilon_n\tilde\epsilon_m}\propto(\tilde\epsilon_{n+1}-\tilde\epsilon_n)=G_n/\tilde\sigma(\tilde\epsilon)$, so it is proportional to the dimension of the species that it acts upon, in accordance with the ansatz (\ref{alpha_nd}) \cite{JPhysA.40.F1013.2007.Anghel,EPL.90.10006.2010.Anghel} and with the general FES rules of Ref. \cite{EPL.87.60009.2009.Anghel}.

If the function $[\delta\tilde\epsilon_{n}/\delta\tilde\rho(\tilde\epsilon_m)] \tilde\sigma(\tilde\epsilon_{n})$ is singular in some point, $\tilde\epsilon_n$, then $\alpha_{\tilde\epsilon_n\tilde\epsilon_m}$ is not proportional to the dimension of the species $G_n$ and
\begin{eqnarray}
  \delta G_n &=& \delta N_m\left[\frac{\delta\bi_{n+1}}{\delta\rho(\tilde\epsilon_m)}\sigma(\bi_{n+1})- \frac{\delta\bi_{n}}{\delta\rho(\tilde\epsilon_m)}\sigma(\bi_{n})\right] \nonumber \\
  &=& -\alpha^{(s)}_{\tilde\epsilon_n\tilde\epsilon_m}\delta N_m . \label{def_alpha_sg}
\end{eqnarray}
The parameters $\alpha^{(s)}_{\tilde\epsilon_n\tilde\epsilon_m}$ obey the rules of Ref. \cite{EPL.87.60009.2009.Anghel} and in addition might still obey the ansatz (\ref{alpha_nd}) proposed in the Introduction. Such a case was discussed in Ref. \cite{PhysLettA.372.5745.2008.Anghel,PhysLettA.376.892.2012.Anghel}, in relation to the Fermi liquid theory. 

Once the FES parameters are known, the thermodynamics follows according to the formalism outlined in the Introduction. 

\section{Conclusions} 

I showed that if a system of interacting particles is described as a gas of quasiparticles of energies, $\tilde\epsilon(\{n_i\})$, such that $E(\{n_i\}) = \sum_i n_i\tilde\epsilon_i$ (condition \ref{cond_entot}), then the gas of quasiparticles may be viewed as an ideal gas that obeys fractional exclusion statistics (FES). If the condition (\ref{cond_entot}) is not satisfied, as it happens with the Landau's quasiparticles in the in the Fermi liquid theory, then the quasiparticle gas cannot be used for the calculation of the thermodynamic properties of the original interacting particle gas, except in some trivial cases (see \cite{arXiv:1204.4064.Anghel}). 

The general method for calculating the FES parameters of the quasiparticle gas is also given. 

\section*{Acknowledgements}

The work was supported by the Romanian National Authority for Scientific Research projects PN-II-ID-PCE-2011-3-0960 and PN09370102/2009. The travel support from the Romania-JINR Dubna collaboration project Titeica-Markov and project N4063 are gratefully acknowledged. 

% \bibliography{/home/dragos/general}
%\bibliographystyle{unsrt}

\end{document}